\documentclass[preprint,12pt]{elsarticle}
\usepackage{lineno,hyperref}
\modulolinenumbers[5]
\usepackage{amsmath,amssymb}    
\usepackage{graphicx}    

\journal{Physica A}
\usepackage{algorithm}  
\usepackage{algorithmicx}  
\usepackage{algpseudocode}  

\usepackage{amsmath}
\usepackage{amsfonts}

\algnewcommand\algorithmicforeach{\textbf{for each}}
\algdef{S}[FOR]{ForEach}[1]{\algorithmicforeach\ #1\ \algorithmicdo}
\usepackage{caption}
\RequirePackage{tikz}
\RequirePackage{zref-abspage}
\RequirePackage{zref-user}
\RequirePackage{tikz}
\RequirePackage{atbegshi}
\usetikzlibrary{calc}
\RequirePackage{tikzpagenodes}
\RequirePackage{etoolbox}

\makeatletter
\newcommand*\ALG@lastblockb{b}
\newcommand*\ALG@lastblocke{e}
\apptocmd{\ALG@beginblock}{%
	\ifx\ALG@lastblock\ALG@lastblockb
	\ifnum\theALG@nested>1\relax\expandafter\@firstoftwo\else\expandafter\@secondoftwo\fi{\ALG@tikzborder}{}%
	\fi
	\let\ALG@lastblock\ALG@lastblockb%
}{}{\errmessage{failed to patch}}

\pretocmd{\ALG@endblock}{%
	\ifx\ALG@lastblock\ALG@lastblocke
	\addtocounter{ALG@nested}{1}%
	\addtolength\ALG@tlm{\csname ALG@ind@\theALG@nested\endcsname}%
	\ifnum\theALG@nested>1\relax\expandafter\@firstoftwo\else\expandafter\@secondoftwo\fi{\endALG@tikzborder}{}%
	\addtolength\ALG@tlm{-\csname ALG@ind@\theALG@nested\endcsname}%
	\addtocounter{ALG@nested}{-1}%
	\fi
	\let\ALG@lastblock\ALG@lastblocke%
}{}{\errmessage{failed to patch}}

\tikzset{ALG@tikzborder/.style={line width=0.5pt,black}}

\newcommand*\currenttextarea{current page text area}

\newcommand*{\updatecurrenttextarea}{%
	\if@twocolumn
	\if@firstcolumn
	\renewcommand*{\currenttextarea}{current page column 1 area}%
	\else
	\renewcommand*{\currenttextarea}{current page column 2 area}%
	\fi
	\else
	\renewcommand*\currenttextarea{current page text area}%
	\fi
}

\newcounter{ALG@tikzborder}
\newcounter{ALG@totaltikzborder}
\newenvironment{ALG@tikzborder}[1][]{%
	\ifx&#1&\else
	\tikzset{ALG@tikzborder/.style={#1}}%
	\fi
	\stepcounter{ALG@totaltikzborder}%
	\expandafter\edef\csname ALG@ind@border@\theALG@nested\endcsname{\theALG@totaltikzborder}%
	\setcounter{ALG@tikzborder}{\csname ALG@ind@border@\theALG@nested\endcsname}%
	\tikz[overlay,remember picture] \coordinate (ALG@tikzborder-\theALG@tikzborder);
	\zlabel{ALG@tikzborder-begin-\theALG@tikzborder}%
	\ifnum\zref@extract{ALG@tikzborder-begin-\theALG@tikzborder}{abspage}=\zref@extract{ALG@tikzborder-end-\theALG@tikzborder}{abspage} \else
	\updatecurrenttextarea
	\ALG@drawvline{[shift={(0pt,.5\ht\strutbox)}]ALG@tikzborder-\theALG@tikzborder}{\currenttextarea.south east}{\ALG@thistlm}%
	\newcounter{ALG@tikzborderpages\theALG@tikzborder}%
	\setcounter{ALG@tikzborderpages\theALG@tikzborder}{\numexpr-\zref@extract{ALG@tikzborder-begin-\theALG@tikzborder}{abspage}+\zref@extract{ALG@tikzborder-end-\theALG@tikzborder}{abspage}}%
	\ifnum\value{ALG@tikzborderpages\theALG@tikzborder}>1
	\edef\nextcmd{\noexpand\AtBeginShipoutNext{\noexpand\ALG@tikzborderpage{\theALG@tikzborder}{\the\ALG@thistlm}}}
	\nextcmd
	\fi
	\fi
}{%
\setcounter{ALG@tikzborder}{\csname ALG@ind@border@\theALG@nested\endcsname}%
\tikz[overlay,remember picture] \coordinate (ALG@tikzborder-end-\theALG@tikzborder);
\zlabel{ALG@tikzborder-end-\theALG@tikzborder}%
\updatecurrenttextarea
\ifnum\zref@extract{ALG@tikzborder-begin-\theALG@tikzborder}{abspage}=\zref@extract{ALG@tikzborder-end-\theALG@tikzborder}{abspage}\relax
\ALG@drawvline{[shift={(0pt,.5\ht\strutbox)}]ALG@tikzborder-\theALG@tikzborder}{ALG@tikzborder-end-\theALG@tikzborder}{\ALG@thistlm}%
\else
\ALG@drawvline{\currenttextarea.north west}{ALG@tikzborder-end-\theALG@tikzborder}{\ALG@thistlm}%
\fi
}

\newcommand*{\ALG@drawvline}[3]{
	\begin{tikzpicture}[overlay,remember picture]
	\draw [ALG@tikzborder]
	let \p0 = (\currenttextarea.north west), \p1=(#1), \p2 = (#2)
	in
	(#3+\fboxsep+14\pgflinewidth+\x0,\y1+\fboxsep+.5\pgflinewidth)
	--
	(#3+\fboxsep+14\pgflinewidth+\x0,\y2-\fboxsep-.5\pgflinewidth)
	;
	\end{tikzpicture}%
}

\newcommand{\ALG@tikzborderpage}[2]{
	\updatecurrenttextarea
	\setcounter{ALG@tikzborder}{#1}%
	\ALG@drawvline{\currenttextarea.north west}{\currenttextarea.south east}{#2}%
	\addtocounter{ALG@tikzborderpages\theALG@tikzborder}{-1}%
	\ifnum\value{ALG@tikzborderpages\theALG@tikzborder}>1
	\AtBeginShipoutNext{\ALG@tikzborderpage{#1}{#2}}%
	\fi
	\vspace{-0.5\baselineskip}
}

\def\ALG@tikzbordertext{\the\ALG@tlm}
\renewcommand{\ALG@beginalgorithmic}{\scriptsize}
\algrenewcommand\alglinenumber[1]{\tiny #1:}
\makeatother

\begin{document}

\begin{frontmatter}

\title{A new fast algorithm for reproducing complex networks with community structure}

\author{Mateusz Kowalczyk, Piotr Fronczak, Agata Fronczak}
\address{Faculty of Physics, Warsaw University of Technology, Koszykowa 75, 00-662 Warsaw, Poland}
\ead{fronczak@if.pw.edu.pl}

\begin{abstract}
In this paper we introduce a new algorithm allowing for generation of networks with heterogeneity of both node degrees and community sizes. The quality and efficiency of the algorithm is analyzed and compared to the other, so far the most popular algorithm which was proposed by Lancichinetti et al. We discuss the advantages and shortcomings of both algorithms indicating the areas of their potential application.
\end{abstract}


\end{frontmatter}

\linenumbers

\section{Introduction}

The community structure is considered to be, next to the small-world effect and scale-free degree distribution, one of the most important topological properties of real networks. By the community (also called cluster, module, or block) in a network we understand a group of nodes more densely connected to each other than to nodes outside the group. For example,
in social networks, communities correspond to groups of people sharing the same interests \cite{Girvan2002}, in Internet, they consist of the sets of web pages on the same topic \cite{Flake}, while in cellular and metabolic networks, communities are functional modules of interacting proteins \cite{Holme2003}.

In the science of complex networks, community detection has become one of the most dominant research topics over the last decade. As a consequence, a large number of algorithms have been proposed for the analysis of community structure in network \cite{PhysRep2010Fortunato, NaturePhys2012Newman,PRE2011Karrer, palla2005uncovering}. To evaluate these algorithms effectively, synthetic networks with a well-defined community structure (benchmarks) had to be proposed. The advantage of such models is that, unlike in real networks, one can easily vary the model parameters and compare the recovered community structure with the predefined one. 

One of the first models of networks with community structure, with a long tradition of study in the social sciences and computer science \cite{SocNet1983Holland, SocNet1992Faust, SocNet1992Anderson, JClass1997Snijders, JMach2008Airoldi, JMach2009Goldenberg}, is the so-called
blockmodel. In its classical version \cite{SocNet1983Holland}, each of $N$ nodes is assigned to one of $K$ blocks (communities) of equal size, and undirected edges are independently drawn between
pairs of nodes with probabilities that are a function only of the group membership of the nodes. Unfortunately, the Poisson-like degree distribution makes this model unsuitable for the further analysis, since most of real networks exhibit power laws in their degree distributions. 

Lancichinetti et al. \cite{PRE2009Lancichinetti} proposed an efficient numerical construction procedure for benchmark graphs that is free of this defect. The method accounts for the heterogeneity in the distributions of node degrees as well as community sizes. Its efficiency has been tested and proved in typical cases, however, further in this paper we show that in a certain range of parameters efficiency of the algorithm drops significantly. Moreover, the complexity of the proposed procedure does not allow for the analytic tractability.

In opposite to Lancichinetti's procedure, Fronczak et al. \cite{PRE2013Fronczak} provided an exponential random
graph formulation \cite{PRE2004Park, PRE2006Fronczak, inbookNewman, Rev2013Shalizi, Essay2012Fronczak} for blockmodel that is solvable for its parameter values in closed forms. Two kinds of the network structural Hamiltonians have been considered: the first one corresponding to the classical blockmodel, and the second one corresponding to its degree-corrected version. In both cases, a number of analytical predictions about various
network properties was given. In particular, it was shown that in the degree-corrected blockmodel, node degrees display an interesting scaling property, that is similar to the scaling feature of the node degrees in fractal (self-similar) real-world networks. Unfortunately, the method is computationally inefficient since it is based on Markov chain Monte Carlo algorithm.

In this contribution we propose a simple, analytically tractable, and fast algorithm for generation of networks with community structure and  heterogeneity of both node degrees and community sizes. The method allows to generate, in a reasonable time, networks that are orders of magnitude larger than those generated by the previous approaches. It also allows for closed-form parameter solutions. 

In outline, the paper is as follows. First, we introduce a new method (KA; the meaning of this abbreviation is "Kowalczyk's et al. algorithm") for generating clustered networks and derive their main properties. Next, we review Lancichinetti's algorithm (LA). We describe its sub-procedures and their time complexity. This allows us to point the range of parameters for which the algorithm efficiency drastically drops down. Finally, we discuss all the major pros and cons of the both approaches. In the appendix, we provide detailed listings of the both algorithms.

\section{Derivation of the new algorithm}
In this section, we present a simple algorithm to generate networks with community structure, which, despite its simplicity, has not been considered, at least to our knowledge, in previous studies. The algorithm is an extension of the model for generating uncorrelated networks with a given sequence of expected degrees $\lbrace \langle k_1\rangle, \langle k_2\rangle,\dots\langle k_N\rangle\rbrace$ (see eg. Eq.~(15) in~\cite{EPJB2004Boguna} and Eq.~(48) in~\cite{PRE2006Fronczak}). In such a prototype network, there is at most one link between any pair of nodes, and there are no self-loops connecting nodes to themselves. If $a_{ij}$ is an entry of the adjacency matrix underlying the network, and $a_{ij}\in\lbrace 0, 1\rbrace$, where $a_{ij}=a_{ji}$ and $a_{ii} = 0$, then the expected value of the entry, $\langle a_{ij}\rangle$, can be expressed in terms of the probability, $p_{ij}$, that the vertices $i$ and $j$ are connected, namely
\begin{equation}\label{apf1}
\langle a_{ij}\rangle=1\cdotp p_{ij}+0\cdotp (1-p_{ij})=p_{ij}.
\end{equation}
Simultaneously, given the expected node degrees, the average number of connections, which obviously can not be greater than one, may be estimated as the expected number of successes in $\langle k_i\rangle$ attempts of $i$ to connect to $j$, where the probability of success for one trial is $\langle k_j\rangle(\sum_{j\neq i} \langle k_j\rangle)^{-1}$, i.e.
\begin{equation}\label{apf2}
\langle a_{ij}\rangle=\langle k_i\rangle\frac{\langle k_j\rangle}{\sum_{j=1}^N \langle k_j\rangle-\langle k_i\rangle}\simeq \frac{\langle k_i\rangle \langle k_j\rangle}{\langle k\rangle N}.
\end{equation}
By comparing Eqs.~(\ref{apf1}) and ~(\ref{apf2}), one gets a simple expression for the probability of a connection:
\begin{equation}\label{pij0}
p_{ij}=\frac{\langle k_i\rangle \langle k_j\rangle}{\langle k\rangle N}.
\end{equation}

In analogy to the above derivation, in networks with community structure, one can write similar relations for the probabilities $p^{int}_{ij}$ and $p^{ext}_{ij}$, that there is an internal or external connection between two nodes, $i$ and $j$, belonging to the same or to different communities. If it is not clear, let us explain that internal connections are those that are between nodes belonging to the same community. Accordingly, the external connections are those that are between nodes belonging to different clusters. 

Thus, let $\langle k_{i,r}^{int}\rangle$ represent the expected internal degree of a node $i$ belonging to the $r-$th community. Correspondingly, let $\langle k_{i,r}^{ext}\rangle$ be the expected number of its external connections. 
Then:
\begin{equation}\label{pint1}
p^{int}_{ij}=\langle k^{int}_{i,r}\rangle\frac{\langle k^{int}_{j,r}\rangle}{\sum_{j=1}^{c_r}\langle k^{int}_{j,r}\rangle-\langle k^{int}_{i,r}\rangle}\simeq\frac{\langle k^{int}_{i,r}\rangle\langle k^{int}_{j,r}\rangle}{2\langle E^{int}_r\rangle},
\end{equation}
and
\begin{equation}\label{pext1}
p^{ext}_{ij}=\langle k^{ext}_{i,r}\rangle \frac{\langle k^{ext}_{j,s}\rangle}{\sum_{s\neq r}\sum_{j=1}^{c_s}\langle k^{ext}_{j,s}\rangle}\simeq\frac{\langle k^{ext}_{i,r}\rangle \langle k^{ext}_{j,s}\rangle}{2\langle E^{ext}\rangle},
\end{equation}
where $c_r$ is the size of the $r-$th cluster, $\langle E^{int}_r\rangle$ is the expected number of internal links within $r$, and $\langle E^{ext}\rangle$ is the number of external links in the whole network. 

Now, let the mixing parameter, $\mu$, describe a share of links which connect each node with nodes belonging to other clusters, i.e.
\begin{equation}\label{as1}
\langle k^{ext}_{i,r}\rangle=\mu\langle k_{i,r}\rangle,
\end{equation}
and
\begin{equation}\label{as2}
\langle k^{int}_{i,r}\rangle=(1-\mu)\langle k_{i,r}\rangle,
\end{equation}
where
\begin{equation}\label{as3}
\langle k_{i,r}\rangle=\langle k_{i,r}^{int}\rangle+\langle k_{i,r}^{ext}\rangle,
\end{equation} 
is the expected total degree of the node $i$, which belongs to the cluster $r$. Using  Eqs.~(\ref{as1})-(\ref{as3}) the connection probabilities, Eqs.~(\ref{pint1}) and (\ref{pext1}), can be rewritten as follows:
\begin{equation}\label{pint2}
p^{int}_{ij}=\frac{(1\!-\!\mu)\langle k_{i,r}\rangle\;(1\!-\!\mu)\langle k_{j,r}\rangle}{(1\!-\!\mu)\sum_{j=1}^{c_r}\langle k_{j,r} \rangle}=\frac{(1\!-\!\mu)\langle k_{i,r}\rangle \langle k_{j,r}\rangle}{\langle k\rangle c_r},
\end{equation}
and
\begin{equation}\label{pext2}
p^{ext}_{ij}=\frac{\mu\langle k_{i,r}\rangle\;\mu\langle k_{j,r}\rangle}{\mu\sum_{j=1}^{N}\langle k_{j,r}\rangle}=\frac{\mu\langle k_{i,r}\rangle\langle k_{j,r}\rangle}{\langle k\rangle N},
\end{equation}
where it has been assumed that the average degree of the nodes within each community is the same as the average degree averaged across the whole network, i.e.
\begin{equation}\label{meank}
\langle k\rangle=\frac{1}{c_r}\sum_{j=1}^{c_r}\langle k_{j,r}\rangle= \frac{1}{N}\sum_{r=1}^n\sum_{j=1}^{c_r}\langle k_{j,r}\rangle,
\end{equation}
where $n$ is the number of clusters.

Having the probabilities $p_{ij}^{int}$ and $p_{ij}^{ext}$ derived, one can generate networks with community structure using the following algorithm:
\begin{enumerate}
	\item For each node $v$ draw an expected degree $\langle k_v\rangle$ from a power distribution $P_{\gamma}(k)\sim k^{-\gamma}$.
	
	\item Generate $n$ clusters with sizes $c_r$ drawn from a power distribution $P_{\beta}(c)\sim c^{-\beta}$. Assign each created cluster to $c_r$ consecutive nodes. The sum of all cluster sizes should not be smaller than the number $N$ of nodes in the network.
	
	\item For each pair of nodes $(i,j)$ add a link with the probabilities given by Eqs.~(\ref{pint2}) or~(\ref{pext2}) depending on whether or not the two nodes share the same cluster. 
\end{enumerate} 
The algorithm is listed in detail as the \textbf{Algorithm~1} in the Appendix.

Before we discuss the quality and efficiency of the presented algorithm we would like to restate the algorithm LA, which was provided by Lancichinetti et al. in Ref.~\cite{PRE2009Lancichinetti}. Nowadays, LA is one of the most frequently cited method for generating clustered network in the literature. Having the both algorithms presented we will able to compare their advantages and shortcomings which will give one a reference point to decide by himself which algorithm would fit better to the specified needs.

\section{The algorithm introduced by Lancichinetti et al.}\label{SectLanci}

Here we restate the LA algorithm and discuss some of the implementation issues that significantly impact its performance. We follow the same notation as in the original article \cite{PRE2009Lancichinetti}. In particular, we assume that the node degrees are drawn from a power-law distribution with the exponent $\gamma$, the community sizes are drawn from a power-law distribution with the exponent $\beta$, the number of nodes is $N$, the minimal, average, and the maximal degree are: $k_{min}$,$\langle k\rangle$, and $k_{max}$, respectively. Furthermore, the mixing parameter $\mu$, as in previous section, describes the share of links that connect each node with nodes belonging to other communities.

The algorithm comprises of several steps that are listed as \textbf{Algorithm 2} in the Appendix.
\begin{enumerate}
	\item For each node $v$ draw a degree $k_v$ from the power-law distribution $P_{\gamma}(k)\sim k^{-\gamma}$.
	
	\item Assign an expected internal degree $\langle k^{int}_v\rangle$ to each node $v$ according to relation $\langle k^{int}_v\rangle=(1-\mu) k_v$, cf.~Eq.~(\ref{as2}). Please note, that, in opposite to the total node degrees, the internal degrees obtained in numerical simulations, $k^{int}_v$, may differ from the expected values, $\langle k^{int}_v\rangle$; the latter can only be realized in average. 
	
	\item Create an initial network using the so-called configuration model \cite{1995MolloyRand}. In this model, at the beginning, exactly $ k_v$ "stubs" or half-edges emanate from each node $v$. Then, the network is constructed by choosing a uniformly random matching on these degree “stubs”. It is worth to note, than the obtained networks can contain self-loops and multi-edges (i.e. they are multigraphs). These represent usually a tiny fraction of	all edges, and one can just discard or collapse them, however for $\gamma<3$ this operation can lead to the so-called structural correlations. 
	
	\item Generate empty clusters with capacities drawn from the power-law distribution $P_{\beta}(c)\sim c^{-\beta}$. The sum of all capacities should not be smaller than the number $N$ of nodes in the network.
	
	\item Assign nodes to clusters. Initially empty clusters are successively filled by the nodes under assumption that the internal degree of the inserted node can not exceed the cluster capacity. If the cluster is full (i.e. when its size equals its capacity), then before inserting a new node, one of the nodes previously assigned to this cluster is removed. This step, as the most affecting the performance of the algorithm is described in detail as \textbf{Algorithm 3} in the Appendix.
	
	\item Perform $N/n$ steps ($n$ is the number of clusters) of the optimization process that tries to minimize deviation between the actual internal degree, $k_v^{int}$, and the expected one, $\langle k_v^{int}\rangle$, namely
	\begin{equation}
	\sigma^2=\sum_{v} \left(\langle k^{int}_v\rangle-k^{int}_v\right)^2. 
	\end{equation} 
	During each step the network configuration is updated via the link rewiring process, which preserves	the degree of each node and affects internal degrees only. 
\end{enumerate} 

\begin{figure}[t]
	\begin{center}
		\includegraphics[width=\textwidth]{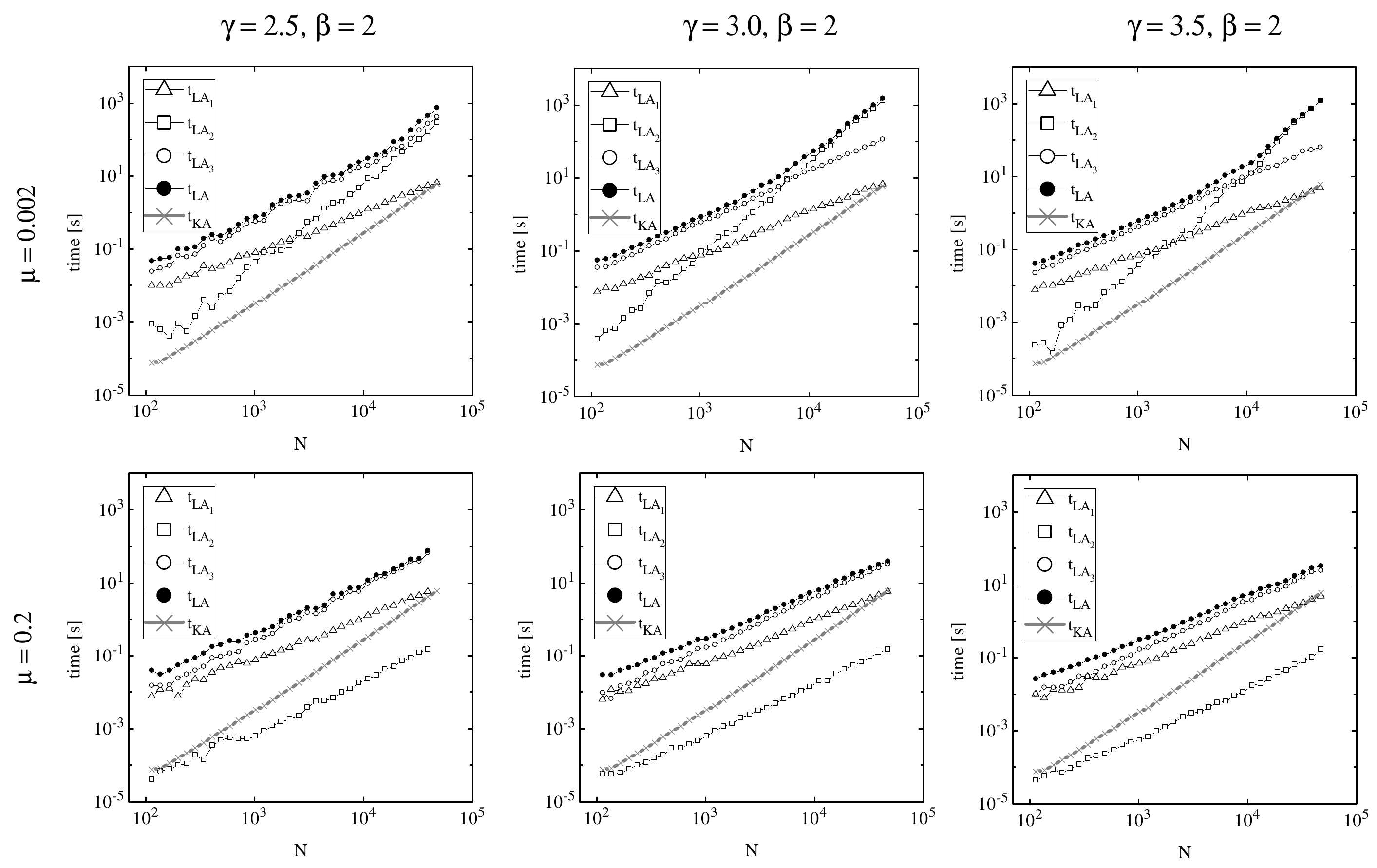}
	\end{center}
	\caption{\label{fig1} Comparison of the total execution time, $t_{KA}$, for the KA algorithm with the total execution time, $t_{LA}$, and the partial times $t_{LA_1}$, $t_{LA_2}$, and $t_{LA_3}$ corresponding to the specified sub-procedures of the LA algorithm. The figure presents data averaged over  
	10 realizations of networks with $\beta=2$, $\langle k\rangle=16$, $k_{max}=\sqrt{\langle k\rangle N}$, and $k_{min}$ given by the normalization condition $k_{min}=\langle k\rangle (\gamma-2)/(\gamma-1)$, and different settings of the parameters $\gamma$ and $\mu$.}
\end{figure}

\section{Comparative analysis of the two algorithms}

\begin{figure}[t]
	\begin{center}
		\includegraphics[width=0.95\textwidth]{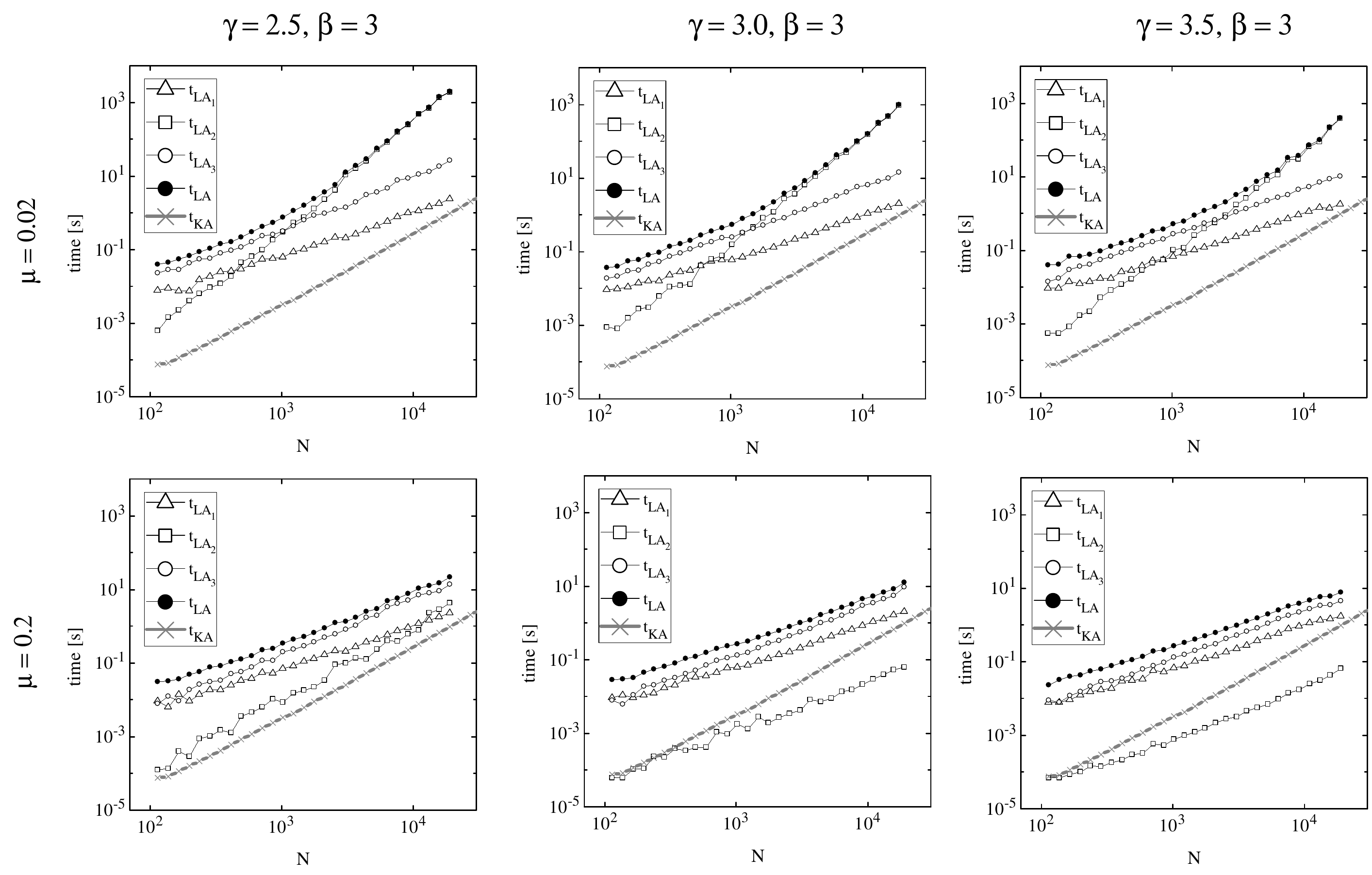}
	\end{center}
	\caption{\label{fig2} The execution time $t_{LA}$ of the LA algorithm, times $t_{LA_1}$, $t_{LA_2}$, and $t_{LA_3}$ of its sub-procedures, compared to the execution time $t_{KA}$ of the KA algorithm for $\beta=3$ and different sets of the parameters $\gamma$ and $\mu$. All the presented results are obtained similarly to those in Fig.~\ref{fig1}.}
\end{figure}

The complexity $\mathcal{O}(N^2)$ of the KA algorithm is obvious due to the iteration over $\binom{N}{2}$ pairs of nodes which can be optionally connected. This complexity does not depend on any other parameter of the model. The execution time $t_{KA}$ of this algorithm for different settings of the parameters $\gamma$, $\beta$, and $\mu$ is presented in Figs.~\ref{fig1} and~\ref{fig2}. 

The efficiency of the LA algorithm has been partially analyzed in \cite{PRE2009Lancichinetti}. The authors state therein that the procedure allows one to build fairly large networks (up to $10^5$-–$10^6$ nodes) in a reasonable time. Extracting data from Fig.~2 in Ref.~\cite{PRE2009Lancichinetti}, one can actually draw such a conclusion. However, as we will show later in this section, the time needed to build such large networks may vary from 30 minutes to 20 days (on a 2.6 GHz Intel Core i5) depending on the choice of parameters $\gamma$, $\beta$, and $\mu$.

To show this, we have analyzed execution times $t_{LA_1}$, $t_{LA_2}$, and $t_{LA_3}$, of the three main sub-procedures comprising the above algorithm (corresponding to the steps 3, 5, and 6 of the construction procedure, which is described in Sect.~\ref{SectLanci}). We have omitted the analysis of other sub-procedures, since they have no visible impact on the total execution time $t_{LA}$.

Regarding the time $t_{LA_1}$ needed to build the configuration model one can estimate its complexity as $\mathcal{O}(N)$. This scaling results from the number of "stubs" that have to be connected, which is twice a number of links $E=\langle k\rangle N$. The complexity $\mathcal{O}(N)$ of the time $t_{LA_3}$ is simply due to the execution of $N/n$ iterations in the sub-procedure $6$. Both these predictions have been confirmed experimentally for different sets of the parameters $\gamma$, $\beta$, and $\mu$ (see Fig.~\ref{fig1} and~\ref{fig2}). 

The most interesting part of the LA algorithm takes place during the assignment of nodes to clusters. In the best case, when each node is assigned to its cluster without hindrance, the complexity of the time $t_{LA_2}$ is simply linear with the system size, $N$. This usually happens when the clusters are large enough to include any node regardless of its expected internal degree. Such a situation can occur in two different ways. First, for sufficiently small expected internal degrees, i.e. for $\mu\rightarrow 1$, what corresponds to fuzzy communities, and second, when all cluster capacities are larger than $k_{max}$, what corresponds to the network consisted of only several large communities.

\begin{figure}[t]
	\begin{center}
		\includegraphics[width=0.95\textwidth]{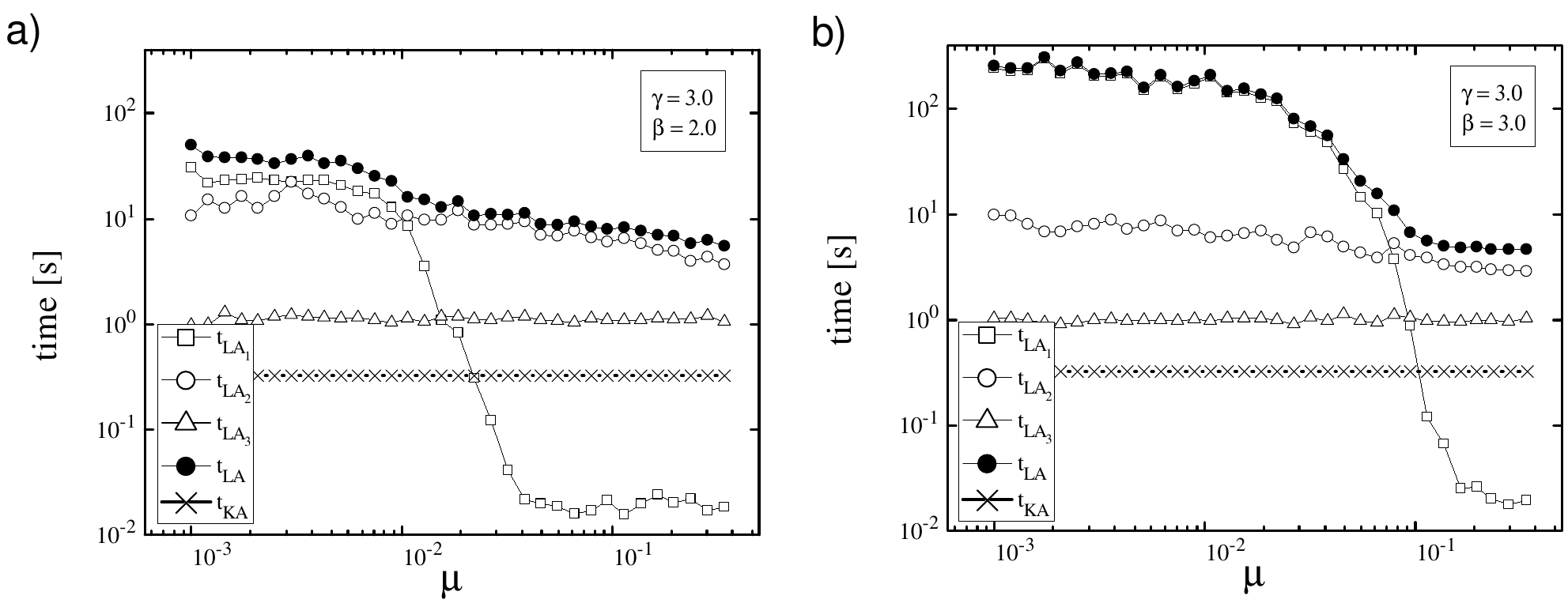}
	\end{center}
	\caption{\label{fig3} The execution time $t_{LA}$ of the LA algorithm, times $t_{LA_1}$, $t_{LA_2}$, and $t_{LA_3}$ of its specified sub-procedures as compared to the execution time $t_{KA}$ of the KA algorithm for $N=10000$ and different settings of the parameters $\gamma$ and $\beta$. All the presented results are obtained similarly to those in Fig.~\ref{fig1}.}
\end{figure}

\begin{figure}[t]
	\begin{center}
		\includegraphics[width=0.55\textwidth]{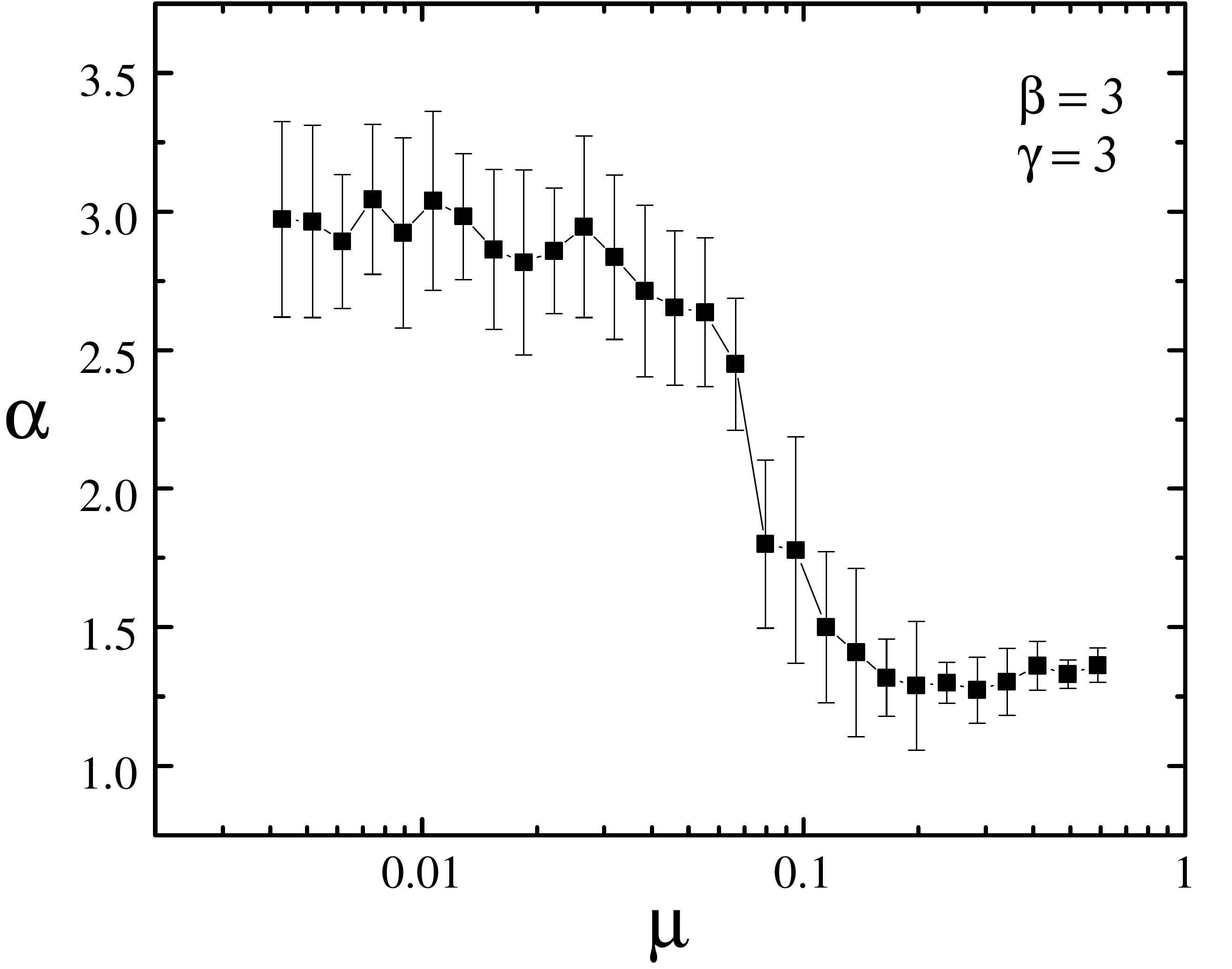}
	\end{center}
	\caption{\label{fig4} Dependence of the scaling exponent $\alpha$ in the relation $t_{LA_2}\sim N^{\alpha}$ on the mixing parameter $\mu$. The complexity of the time $t_{LA_2}$ ranges between $1$ (overlapping communities) and $3$ (clearly separated communities).}
\end{figure}

In the worst case, when all the cluster capacities are comparable with the expected internal degrees of nodes, the algorithm iterates $3N$ times trying to find an appropriate cluster for each node (what executes $3N*N$ times in total). If, after those $3N$ trials, there are still unassigned nodes, then the two smallest clusters are merged and the whole process repeats. 

To see how does it work, let us shortly discuss the case of $\gamma\approx\beta$, i.e., when the both, node degrees and cluster capacities, are drawn from the same distribution. Then, the average cluster capacity $\langle c\rangle\approx\langle k\rangle$, and the merging process can repeat $n$ times, where $n$ is the number of clusters $n\approx N/\langle c\rangle\approx N/\langle k\rangle$. Taking all these iterations into account, one can estimate the complexity of the time $t_{LA_2}$ as $\mathcal{O}(N^3)$. As one can see in Fig.~\ref{fig1} and~\ref{fig2}, the time $t_{LA_2}$ becomes a dominant factor for the whole processing time $t_{LA}$ for $\beta=2$ and $\gamma=3$ when $N>10^4$, and, for $\beta=\gamma=3$ when $N>10^3$, i.e. for networks of medium size. Figs.~\ref{fig3} and~\ref{fig4} demonstrate, how the
complexity of $t_{LA_2}$ depends on the mixing parameter $\mu$. This is shown in comparison to the other analysed times. In this figure, one can see the remarkable transition from the linear, $\mathcal{O}(N)$, scaling of $t_{LA}$ for $\mu\rightarrow 1$, to the cubic-like, $\mathcal{O}(N^3)$, regime for $\mu\ll 1$. 

\begin{figure}[t]
	\begin{center}
		\includegraphics[width=0.95\textwidth]{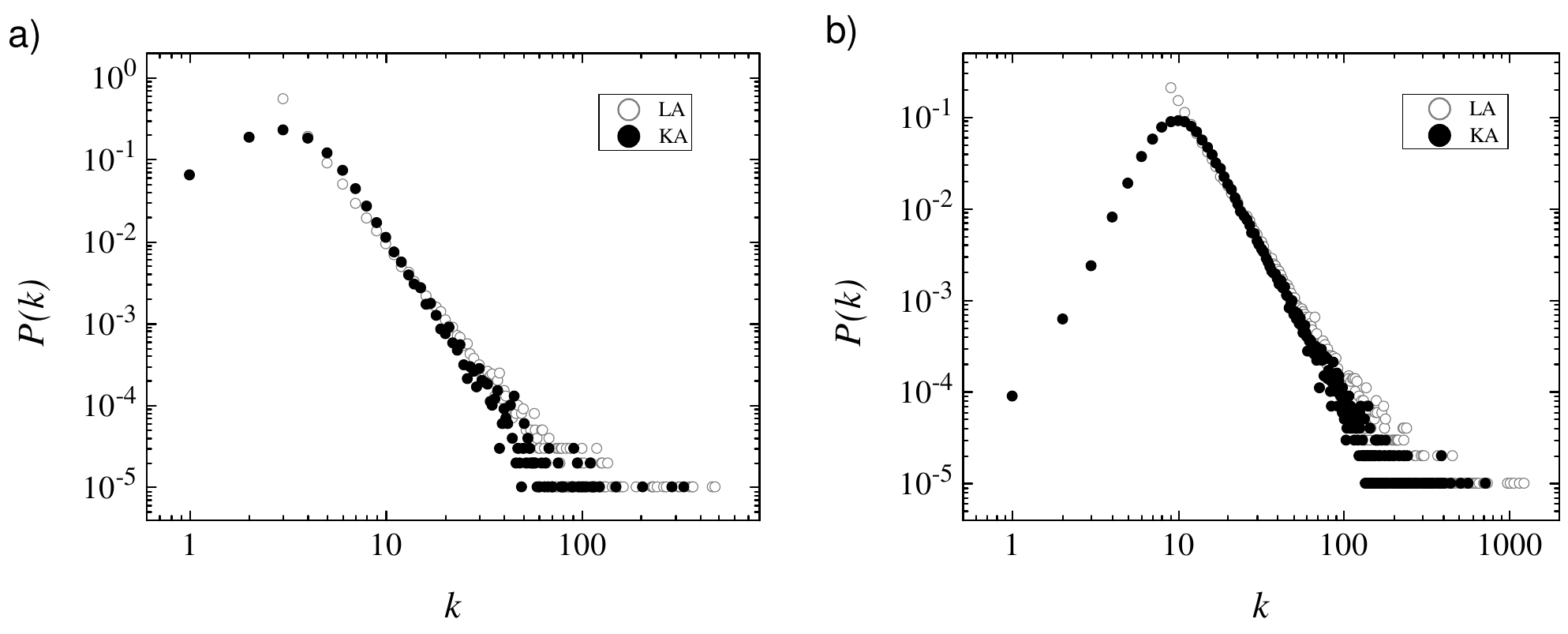}
	\end{center}
	\caption{\label{fig5} Node degree distributions obtained by numerical simulations using both algorithms, KA and LA. The figure presents results for single networks of size $N=100000$, $\gamma=3$, and two different values of the expected node degrees: (a) $\langle k \rangle=4$, and (b) $\langle k \rangle=16$.}
\end{figure}

\begin{figure}[t]
	\begin{center}
		\includegraphics[width=0.95\textwidth]{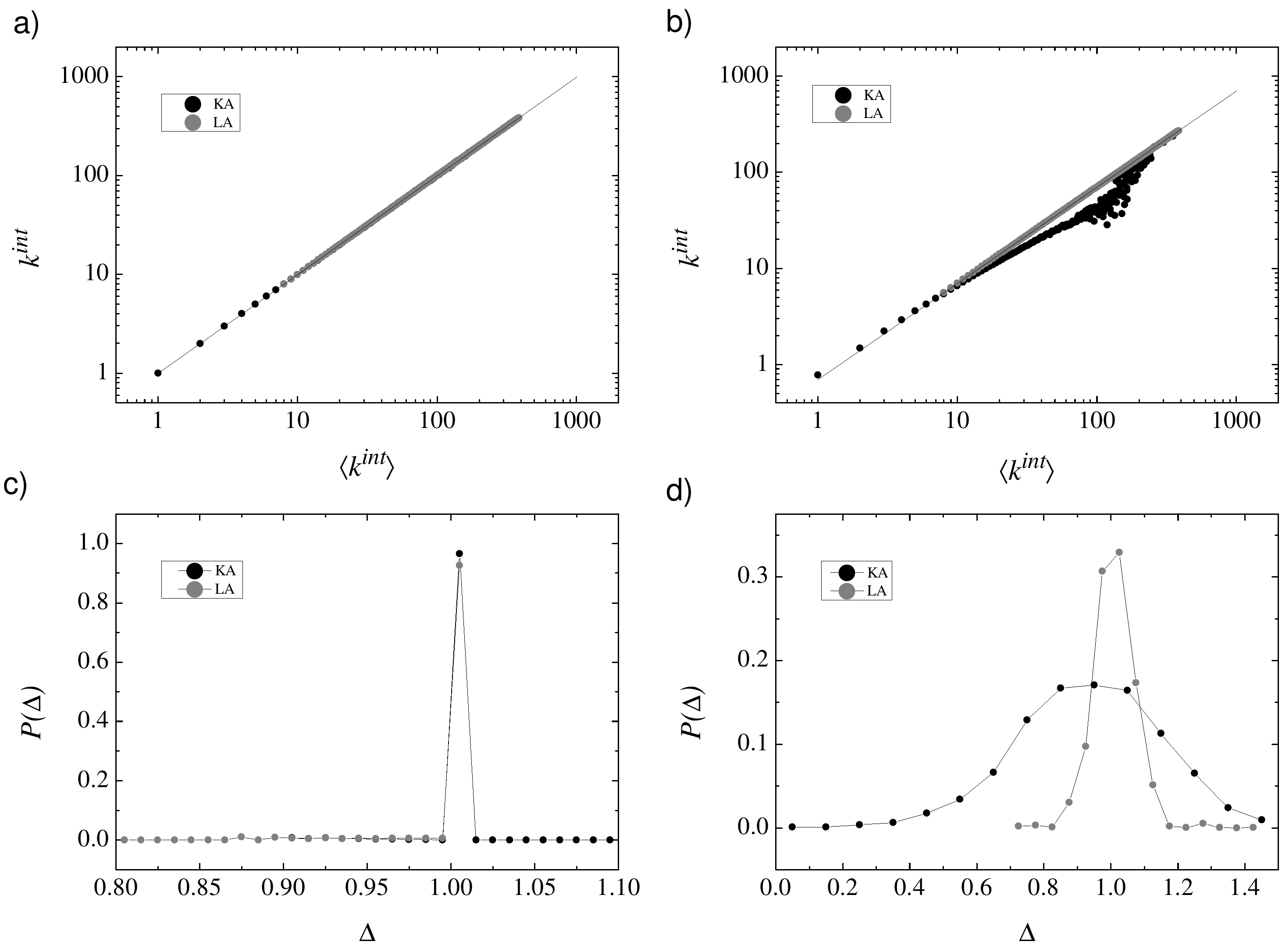}
	\end{center}
	\caption{\label{fig6} Obtained internal node degrees, $k^{int}$ in relation to their expected values, $\langle k^{int}\rangle$, and the distributions $P(\Delta)$ of the deviation $\Delta=k^{int}/\langle k^{int}\rangle$ for $\mu=0.01$ (a,c) and for $\mu=0.3$ (b,d). The results are averaged over 100 realizations of networks with $N=10000$ and $\gamma=\beta=3$.}
\end{figure}

The above findings suggest possible application areas of the described algorithm. It has a potential to generate large networks under assumption that the communities are fuzzy ($\mu\rightarrow 1$) or that there is a small number of sufficiently large clusters.

Comparing the time execution of the both algorithms one can state that for moderate network sizes the KA outperforms the LA by orders of magnitude. Extrapolating straight lines in Fig.~\ref{fig1} and~\ref{fig2} one can estimate that, for the fuzzy communities, $\mu\rightarrow 1$, $t_{KA}$ becomes larger than $t_{LA}$ for $N>10^6$, i.e. for really large networks. In the case of well defined communities, $\mu\ll 1$, the time $t_{KA}$ will never exceed the time $t_{LA}$.

Let us now discuss the quality of the both algorithms. It can be assessed on two levels, namely the level of total node degrees and the level of internal node degrees, see Eq.~(\ref{as3}). On the first level, each node, $i$, of the considered networks is characterized by two parameters: the expected degree, $\langle k_i\rangle$, and the obtained degree, $k_i$. In the case of LA, both these quantities are equal since, after assigning the expected node degrees, one just matches together half-edges emanating from each node. In the case of KA, the probabilistic character of connections between different pairs of nodes leads to asymptotic scale-free networks. The resulting node degree distribution is blurry as compared with the expected one (see Fig.~\ref{fig5}). This is due to the fact that the obtained distribution is a kind of convolution of the expected scale-free distribution and the Poisson distribution \cite{PREBogunahidden, PRE2006bFronczak}. The Poisson-like blur of each node degree is the most perceptible for low degree nodes. For medium and large degrees (hubs) it is almost imperceptible.

The direct consequence of the mentioned blur is the occurrence in KA networks isolated nodes. The number of such zero-degree nodes $N_{k=0}$ strongly depends on the average node degree, $\langle k\rangle$, and it can be significant in sparse networks. For example, $N_{k=0}\approx 0.1 N$ in the KA network shown in Fig.~\ref{fig5}a) for which $\langle k\rangle=4$, and $N_{k=0}\approx 0.00006 N$ in the KA network with $\langle k\rangle=16$, shown in Fig.~\ref{fig5}b). We numerically checked that $N_{k=0}$ decreases exponentially with $\langle k\rangle$.

One also has to keep in mind that in both algorithms, the so-called structural correlations which occur for $\gamma<3$ may lead to discrepancies between $k_i$ and its expected value, $\langle k_i\rangle$. To avoid them, one has to assure that $k_{max}<\sqrt{\langle k\rangle N}$ (cf. Eq.~(16) in~\cite{EPJB2004Boguna}). 

On the second level of the analysis each node of the considered networks can be characterized by the two corresponding parameters: the expected internal degree, $\langle k^{int}_v\rangle$, and the obtained degree, $k^{int}_v$. The both algorithms can achieve the agreement between the both quantities only in average, however the optimization performed in the step 6 of the LA suggests that this algorithm should be much more precise than the KA in this context. The comparison of both algorithms presented in Fig.~\ref{fig6} confirms this statement.

Deviations between $\langle k^{int}_v\rangle$ and $k^{int}_v$ in KA result from the approximation in Eq.~(\ref{pint1}), namely from the fact that we neglect the node degree $\langle k^{int}_{i,r}\rangle$ in the denominator of this expression. Such an approximation is crude when the considered node $i$ is a hub (i.e. for~$\langle k^{int}_{i,r}\rangle\approx k_{max}$) and the sum of the degrees of the rest of the nodes in the cluster is small. The approximation will work much better if the clusters are dense, i.e. for $\mu\ll 1$. This conclusion is confirmed by the differences in the quality of the KA algorithm for $\mu=0.01$ (Fig.~\ref{fig6}a and \ref{fig6}c) and $\mu=0.3$ (Fig.~\ref{fig6}b and \ref{fig6}d). 

\begin{table}\label{tab1}
	\centering
	\begin{tabular}{|c||c|c|}
		\hline
		characteristics	& LA  & KA \\  \hline \hline 
		time efficiency	& $-$ & + \\  \hline
		quality of the generated networks	& + & $-$ \\  \hline
		analytical solution	& $-$ & + \\  \hline
		no zero-degree nodes	& + & $-$  \\ \hline
		simplicity of implementation	& $-$ & +  \\ \hline
	\end{tabular}
	\caption{A comparison of the all discussed characteristics of the both algorithms.}
\end{table}

\section{Conclusions}

Both algorithms have their own advantages and shortcomings. They are gathered in the Table 1. While choosing an adequate algorithm one has to consider a trade off between accuracy, speed and analytical tractability of the algorithms. It is clear, however, that KA is much faster and allows to generate huge networks ($N>10^6$) in a reasonable time. It can be easily described analytically (and probably expanded, e.g. taking into account node-degree correlations and even overlaps of communities). On the other hand, LA is much more precise. The variance between expected and obtained node degrees is strongly reduced thanks to the implemented optimization stage. Finally, the algorithms provided in the Appendix, as well as the source codes for KA \cite{cppKA} and for LA \cite{cppLA}, clearly demonstrate that the former outperforms the later in term of the simplicity of implementation.
\appendix
\section{Algorithms reproducing graphs with community structures}

Here we provide listings of the both algorithms for generating networks with community structures. Due to its complexity, we decided to show only overview of the LA method (\textbf{Algorithm 2}). Lines 7, 8, and 9 in the \textbf{Algorithm 2} are in fact sub-procedures, and we provide the detailed listing of the second of them only (as the \textbf{Algorithm 3}). The reason is that all these sub-procedures are quite complicated and that we discuss only this second one in the paper in a more detailed way. On the contrary, the \textbf{Algorithm 1} presents KA method with all the details.
 
\label{sec:Annex1}
	\begin{algorithm}[H]  
		\caption{KA algorithm reproducing graphs with community structure} 
		\begin{algorithmic}[1]  \tiny
			\Require minimal degree $k_{min}$, maximal degree $k_{max}$, network size $N$, heterogeneity coefficients $\gamma$ and $\beta$, mixing parameter $\mu$ 
			\Ensure graph $\mathbb{G}(\mathbb{V},\mathbb{E})$ and map of nodes into clusters $f$
			\Function {Benchmark}{$k_{min},k_{max},N,\gamma,\beta,\mu$} 
			\State Let $\mathbb{V} =\left\lbrace v_i\mid i=1,2,...,N\right\rbrace $ be a sequence of nodes in a graph
			\State Let $\mathbb{E} =\left\lbrace (u,v)\mid v,w\in\mathbb{V}\right\rbrace $ be a set of edges 
			\ForEach {node $v\in \mathbb{V}$}
			\State draw an expected node degree $\langle k_v\rangle$ from a power distribution $P_{\gamma}(k)$
			\EndFor
			\State $\left\langle k\right\rangle \gets \sum_{v\in \mathbb{V}} \langle k_v\rangle/N$
			\State Let $\mathbb{C} = \left\lbrace c_i\mid i\in\mathbb{N}\right\rbrace$ be a sequence of cluster sizes
			\State Let $f:\mathbb{V}\rightarrow\mathbb{S}$ be a map of $\mathbb{V}$ into the set of clusters $\mathbb{S}$
			\State Let $n \gets 0$ be an initial  number of clusters
			\Repeat
			\State $n \gets n+1$
			\State draw a cluster capacity $c_n$ from a power distribution $P_{\beta}(c)$
			\State $totalcapacity \gets \sum_{i\leq n} c_i$
			\ForEach {node $(v_j\mid totalcapacity-c_n< j\leq totalcapacity)$}
			\State $f(j)=n$
			\EndFor
			\Until{$totalcapacity<N$}
			
			\ForEach {node $v\in \mathbb{V}$}
			\ForEach {node $w\in \mathbb{V}$}
			\If{$f(v)=f(w)$}
			\State $p=(1-\mu) \langle k_v\rangle \langle k_w\rangle/(\left\langle k\right\rangle c_{f(v)})$
			\Else 
			\State $p=\mu \langle k_v\rangle \langle k_w\rangle / (\left\langle k\right\rangle N)$
			\EndIf
			\If{$random<p$}
			\State $k_v \gets k_v+1$
			\State $k_w \gets k_w+1$
			\State $\mathbb{E} \gets \mathbb{E} \cup (v,w)$
			\If{$f(v)=f(w)$}
			\State $k^{int}_v \gets k^{int}_v+1$
			\State $k^{int}_w \gets k^{int}_w+1$
			\EndIf
			\EndIf
			\EndFor
			\EndFor
			\State \Return{$\mathbb{G}(\mathbb{V},\mathbb{E})$ and  $f$}  
			\EndFunction  
		\end{algorithmic}  
	\end{algorithm} 

	\begin{algorithm}[H] 
		\caption{LA algorithm reproducing graphs with community structure} 
		\begin{algorithmic}[1] \tiny
			\Require minimal degree $k_{min}$, maximal degree $k_{max}$, network size $N$, heterogeneity coefficients $\gamma$ and $\beta$, mixing parameter $\mu$ 
			\Ensure graph $\mathbb{G}(\mathbb{V},\mathbb{E})$ and nodes-to-clusters assignment $\mathbb{S}$
			\Function {Benchmark}{$k_{min},k_{max},N,\gamma,\beta,\mu$} 
			\State Let $\mathbb{V} =\left\lbrace v_i\mid i=1,2,...,N\right\rbrace $ be a sequence of nodes in a graph
			\ForEach {node $v\in \mathbb{V}$}
			\State draw a node degree $k_v$ from a power distribution $P_{\gamma}(k)$
			\State assign an expected internal degree $\langle k^{int}_v\rangle \gets (1-\mu)k_v$ to a node $v$
			\EndFor
			\State build a preliminary network $\mathbb{G}(\mathbb{V},\mathbb{E})$ using the configuration model
			\State assign nodes to clusters
			
			\State rewire internal and external connections
			\State \Return{$\mathbb{G}(\mathbb{V},\mathbb{E})$ and nodes-to-clusters assignment $\mathbb{S}$}  
			\EndFunction  
			\end{algorithmic}  
			\end{algorithm} 
			\begin{algorithm}[H] 
			\caption{Procedure that assigns nodes to clusters in LA algorithm} 
			\begin{algorithmic}[1] \tiny
			\Procedure {AssignNodesToClusters}{}
			\State Let $\mathbb{C} = \left\lbrace c_i\mid i\in\mathbb{N}\right\rbrace$ be a sequence of cluster capacities
			\State Let $\mathbb{S}=\left\lbrace S_i \mid i\in\mathbb{N}\right\rbrace$ be a sequence of nodes' sequences
			\State Let $S_i =\left\lbrace s^{(i)}_j\mid j\in\mathbb{N}\right\rbrace $ be a sequence of nodes in cluster $i$
			
			\State Let $n \gets 0$ be an initial  number of clusters
			\Repeat
			\State $n \gets n+1$
			\State draw a cluster capacity $c_n$ from a power distribution $P_{\beta}(c)$
			\State create empty cluster $S_n \gets \O$
			\State $totalcapacity \gets \sum_{i\leq n} c_i$
			\Until{$totalcapacity<N$}
			
			\State Let $\mathbb{Z} \gets \mathbb{V}$ be a sequence of nodes currently unassigned to clusters
			\State $trial \gets 0$
			\While{$\mathbb{Z}\neq \O \ \textbf{and} \ n>1$}
			\State $trial \gets trial+1$
			\ForEach {node $v \in \mathbb{Z}$}
			\State select a random cluster $S_i$ from $\mathbb{S}$
			\If{$\langle k^{int}_v\rangle <c_i$} \Comment{if cluster is large enough}
			\If{$|S_i|=c_i$} \Comment{if cluster is full}
			\State $\mathbb{Z} \gets \mathbb{Z} \cup \left\lbrace s^{(i)}_1 \right\rbrace $ \Comment{move $1^{st}$ node from $S_i$ back into $\mathbb{Z}$}
			\State $S_i \gets S_i \setminus \left\lbrace s^{(i)}_1 \right\rbrace$   
			\EndIf
			\State $S_i \gets S_i \cup \left\lbrace v\right\rbrace $ \Comment{add node $v$ to $S_i$} 
			\State $\mathbb{Z} \gets \mathbb{Z} \setminus \left\lbrace v\right\rbrace $
			\EndIf
			\EndFor
			\If{$trial>3N$}
			\State $trial \gets 0$
			\State merge the two smallest clusters into one cluster
			\State $n \gets n-1$
			\State $\mathbb{Z} \gets \mathbb{V}$ 
			\State $\forall S_i\in \mathbb{S} \ \ S_i \gets \O$ 
			\EndIf
			\EndWhile
			\ForEach {cluster $S_i\in \mathbb{S}$} 
			\If{$\sum_{v\in S_i}\langle k^{int}_v\rangle$ is odd}
			\State change $\langle k^{int}\rangle$ of randomly selected node by 1
			\EndIf
			\EndFor			
			\EndProcedure			
		\end{algorithmic}  
	\end{algorithm} 

\section*{References}

\bibliography{lanci_apf}

\begin{thebibliography}{10}
\expandafter\ifx\csname url\endcsname\relax
  \def\url#1{\texttt{#1}}\fi
\expandafter\ifx\csname urlprefix\endcsname\relax\def\urlprefix{URL }\fi
\expandafter\ifx\csname href\endcsname\relax
  \def\href#1#2{#2} \def\path#1{#1}\fi

\bibitem{Girvan2002}
M.~Girvan, M.~E.~J. Newman, Community structure in social and biological
  networks, Proc. Nat. Acad. Sci. 99~(12) (2002) 7821--7826.

\bibitem{Flake}
G.~W. Flake, S.~Lawrence, C.~L. Giles, F.~M. Coetzee, Self-organization and
  identification of web communities, Computer 35~(3) (2002) 66--71.

\bibitem{Holme2003}
P.~Holme, M.~Huss, H.~Jeong, Subnetwork hierarchies of biochemical pathways,
  Bioinformatics 19~(4) (2003) 532--538.

\bibitem{PhysRep2010Fortunato}
S.~Fortunato, Community detection in graphs, Phys. Rep. 486 (2010) 75--174.

\bibitem{NaturePhys2012Newman}
M.~E.~J. Newman, Communities, modules and large-scale structure in networks,
  Nature Physics 8 (2012) 25--31.

\bibitem{PRE2011Karrer}
B.~Karrer, M.~E.~J. Newman, Stochastic blockmodels and community structure in
  networks, Phys. Rev. E 83 (2011) 016107.

\bibitem{palla2005uncovering}
G.~Palla, I.~Derényi, I.~Farkas, T.~Vicsek, Uncovering the overlapping
  community structure of complex networks in nature and society, Nature 435
  (2005) 814--818.

\bibitem{SocNet1983Holland}
P.~W. Holland, K.~B. Laskey, S.~Leinhardt, Stochastic blockmodels: first steps,
  Soc. Networks 5 (1983) 109--137.

\bibitem{SocNet1992Faust}
K.~Faust, S.~Wasserman, Blockmodels: Interpretation and evaluation, Soc.
  Networks 14 (1992) 5.

\bibitem{SocNet1992Anderson}
C.~J. Anderson, S.~Wasserman, K.~Faust, Building stochastic blockmodels, Soc.
  Networks 14 (1992) 137.

\bibitem{JClass1997Snijders}
T.~A. Snijders, K.~Nowicki, Estimation and prediction for stochastic
  block-structures for graphs with latent block structure, J. Classification 14
  (1997) 75--100.

\bibitem{JMach2008Airoldi}
E.~M. Airoldi, D.~M. Blei, S.~E. Fienberg, X.~P. Xing, Mixed-membership
  stochastic blockmodels, J. Mach. Learn. Res. 9 (2008) 1981--2014.

\bibitem{JMach2009Goldenberg}
A.~Goldenberg, A.~X. Zheng, S.~E. Feinberg, E.~M. Airoldi, A survey of
  statistical network models, Found. Trends Mach. Learn. 2 (2009) 1.

\bibitem{PRE2009Lancichinetti}
A.~Lancichinetti, S.~Fortunato, F.~Radicchi, Benchmark graphs for testing
  community detection algorithms, Phys. Rev. E 78 (2008) 046110.

\bibitem{PRE2013Fronczak}
P.~Fronczak, A.~Fronczak, M.~Bujok, Exponential random graph models for
  networks with community structure, Phys. Rev. E 88 (2013) 032810.

\bibitem{PRE2004Park}
J.~Park, M.~E.~J. Newman, Statistical mechanics of networks, Phys. Rev. E 70
  (2004) 066117.

\bibitem{PRE2006Fronczak}
A.~Fronczak, P.~Fronczak, J.~A. Ho{\l}yst, Fluctuation-dissipation relations in
  complex networks, Phys. Rev. E 73 (2006) 016108.

\bibitem{inbookNewman}
M.~E.~J. Newman, Networks. An Introduction, Oxford University Press, Oxford,
  2010, Ch. 15.2, pp. 565--588.

\bibitem{Rev2013Shalizi}
C.~R. Shalizi, A.~Rinaldo, Consistency under sampling of exponential random
  graph models, arXiv:1111.3054v3 [math.ST] (2011).

\bibitem{Essay2012Fronczak}
A.~Fronczak, Exponential random graph models, arXiv:1210.7828 [physics.soc-ph]
  (2012).

\bibitem{EPJB2004Boguna}
M.~Bogu{\~{n}}{\'a}, R.~Pastor-Satorras, A.~Vespignani, Cut-offs and finite
  size effects in scale-free networks, Eur. Phys. J. B. 38~(2) (2004) 205--209.

\bibitem{1995MolloyRand}
M.~Molloy, B.~Reed, A critical point for random graphs with a given degree
  sequence, Random Structures and Algorithms 6 (1995) 161--179.

\bibitem{PREBogunahidden}
M.~Bogu{\~{n}}{\'a}, R.~Pastor-Satorras, Class of correlated random networks
  with hidden variables, Phys. Rev. E 68 (2003) 036112.

\bibitem{PRE2006bFronczak}
A.~Fronczak, P.~Fronczak, Networks with given two-point correlations: Hidden
  correlations from degree correlations, Phys. Rev. E 74 (2006) 026121.

\bibitem{cppKA}
M.~Kowalczyk, P.~Fronczak, A.~Fronczak, A software package to generate graphs
  by the described algorithm can be downloaded from http://
  if.pw.edu.pl/$\sim$agatka/benchmark.zip.

\bibitem{cppLA}
A.~Lancichinetti, S.~Fortunato, F.~Radicchi, A software package to generate
  graphs by the lancichinetti algorithm can be downloaded from http://
  santo.fortunato.googlepages.com/benchmark.tgz.

\end{thebibliography}

\end{document}